\documentclass[submission,copyright,creativecommons]{eptcs}
\providecommand{\event}{WORDS 2011} 
\usepackage{breakurl}             

\usepackage{amssymb}
\usepackage{amsthm}

\def\lcm{\mbox{\rm lcm}}

\def\Z{\mathbb Z}
\def\N{\mathbb N}

\def\Q{\mathbb Q}

\newtheorem{Th}{Theorem}

\newtheorem{Prop}{Proposition}

\title{Circular words and applications} 
\def\titlerunning{Circular words and applications}

\author{Beno\^{\i}t Rittaud
\institute{Laboratoire Analyse, G\'eom\'etrie et Applications, Institut Galil\'ee, Universit\'e Paris-13\\
99 avenue Jean-Baptiste Cl\'ement, 93~430 Villetaneuse, France}
\email{rittaud@math.univ-paris13.fr}
\and
Laurent Vivier
\institute{Laboratoire de Didactique Andr\'e Revuz, Universit\'e Paris Diderot\\
Site Chevaleret -- Case 7018, 75~205 Paris Cedex 13}
\email{laurent.vivier@univ-orleans.fr}}

\def\authorrunning{B. Rittaud \& L. Vivier}

\begin{document} 

\maketitle 

\begin{abstract} 
We define the notion of {\em circular words}, then consider on such words a constraint derived from the Fibonacci condition.
We give several results on the structure of these circular words, then mention possible applications to various situations:
periodic expansion of numbers in numeration systems, ``$\gcd$-property'' of integer sequences, partition of the prefix
 of the fixed point of the Fibonacci substitution, spanning trees of a wheel. Eventually, we mention some open questions.
\end{abstract}

\bigskip

\noindent
Let $b>1$ be an integer. It is well-known that the expansion in base $b$ of a number $x$ is ultimately periodic iff $x$ is rational. Consider $x$ and $x'\in\Q$, and write $W$ and $W'$ for their periodic part, assuming $|W|=|W'|$ (this latter hypothesis can be fulfilled by considering suitable powers of the minimal period of $x$ and $x'$). There exists a simple way to get the periodic part of $x+x'$: it consists in adding $W$ and $W'$, with the only difference from common addition that the possible carry at the last digit on the left has to be reported on the right of the period. This way of adding finite words on an alphabet of numbers embedd the set of finite words of length $\ell$ with a group structure, the group of {\em (punctured) circular words} of length $\ell$ on the alphabet $\{0,\ldots, b-1\}$. To distinguish circular words from standard words, we write $\widetilde{W}$ for the former ones. Formally, a circular word can be regarded as a biinfinite purely periodic wor
 d with a pointer on one of its letters, or as the ordered set defined by a finite word $W=w_0\ldots w_{n-1}$ and its circular shifts $\sigma(W)$, \ldots, $\sigma^{n-1}(W)$, where $\sigma(w_0\ldots w_{n-1}):=w_{n-1}w_0\ldots w_{n-2}$.

To be able to properly define an addition on circular words, an important point is that the circular word $\widetilde{0^n}$ (made of $n$ zeroes) has to be identified with $\widetilde{(b-1)^n}$ (made of $n$ times the digit $b-1$). For $b$ equal to ten, this corresponds to the classical identification $1.000\ldots=0.999\ldots$. Hence, since the group of circular words of length $n$ is abelian, is generated by $\widetilde{0^{n-1}1}$ and has $b^n-1$ elements, it corresponds to $\Z/(b^n-1)\Z$. 

Now, consider the set of circular words of order $q$. This also defines a finite abelian group, which is the group of periods of all the rational numbers of the form $p/q$ (the identification $\widetilde{W^m}=\widetilde{W}$ for any $m\geq 1$ ensure that all these periods can be supposed of equal length). It is easily seen that such a group is isomorphic to $\Z/q\Z$. One of its elements plays a particular role: the one corresponding to the smallest value (apart from the word containing only $0$s). This one, denoted by $\widetilde{\Pi}$, has the particularity that $2\widetilde{\Pi}$, $3\widetilde{\Pi}$,\ldots, $q\widetilde{\Pi}$ are computed by standard classical addition, that is: no carry at the end of the calculation has to be put in the beginning of the word. In other words, we have $i\widetilde{\Pi}=\widetilde{i\Pi}$ for any $i\leq q$. For example, in base ten, the group $\Z/7\Z$ of circular words of order $q=7$ is made of $\widetilde{\Pi}=\widetilde{142857}$, $2\widetilde
 {\Pi}=\widetilde{285714}$, $3\widetilde{\Pi}=\widetilde{428571}$, $4\widetilde{\Pi}=\widetilde{571428}$, $5\widetilde{\Pi}=\widetilde{714285}$, $6\widetilde{\Pi}=\widetilde{857142}$ and $7\widetilde{\Pi}=\widetilde{999999}=\widetilde{000000}=0\widetilde{\Pi}$.

\section{Admissible circular words}

A first question about circular words is to ask how the previous observations extend when considering constraints on words derived from more general dynamical systems. We focus here on the Fibonacci condition: a circular word $\widetilde{W}:=\widetilde{w_0\ldots w_{n-1}}$ on the alphabet $\{0,1\}$ will be said to be {\em admissible} iff it does not contains the factor $11$, that is: for any $0<i<n$ we have $w_{i-1}w_{i}\neq 11$, and $w_{n-1}w_0\neq 11$. (To be more synthetic, the indices of the letters can be considered modulo $n$, so the previous condition can be rewritten as: $w_{i-1}w_i\neq 11$ for any $i$.)

For $W=w_0\ldots w_{n-1}$ and $W'=w'_0\ldots w'_{n-1}$, $\widetilde{W}+\widetilde{W'}$ is achieved by considering the word made of the letters $w_i+w'_i$ and by using as many times as needed the identities \[\ldots x_{k-2}110x_{k+2}\ldots=\ldots x_{k-2}001x_{k+2}\ldots\]
\noindent (which corresponds to the equality $F_{k-1}+F_{k-2}=F_k$, where $(F_k)_k$ is the Fibonacci sequence with $F_0=1$ and $F_1=2$) and
\[\ldots x_{k-3}0020x_{k+2}\ldots=\ldots x_{k-3}1001x_{k+2}\ldots\]
\noindent (which corresponds to the equality $2F_k=F_{k-2}+F_{k+1}$, which is true for any $k\geq 2$).  Note that we write from left to right and not from right to left.

The circular words $\widetilde{W}$ and $\widetilde{W'}$ are {\em equivalent} iff they belong to the same orbit under the previous transformations. It can be shown that, essentially, any circular word $\widetilde{W}$ on the alphabet $\N$ possesses a unique equivalent admissible circular word, denoted by ${\tilde Z}(\widetilde{W})$. The only exception concerns the orbit of the circular word $\widetilde{1^n}$. First, the orbit of $\widetilde{1^{2\ell+1}}$ contains no admissible circular word. For this reason, in the sequel, we consider only circular words of even length $2\ell$. Second, the circular word $\widetilde{1^{2\ell}}$ is equivalent to two admissible circular words: $\widetilde{(01)^\ell}$ and $\widetilde{(10)^\ell}$. Assuming that these two words are equal is enough to get the following group structure.

\begin{Th}\label{Groupe} For any $\ell\geq 1$, define the set ${\cal G}_\ell^*$ of circular admissible words of length $2\ell$ that contains at least one $1$. Assume the identification $\widetilde{(01)^\ell}=\widetilde{(10)^\ell}$. The set ${\cal G}_\ell^*$ is an abelian group for the previous addition. More precisely, we have

\[{\cal G}_\ell^*\simeq\left\{\begin{array}{ll}

(\Z/d_\ell\Z)\times(\Z/d_\ell\Z) & \mbox{for $\ell$ odd;}\\
(\Z/5d_\ell\Z)\times(\Z/d_\ell\Z) & \mbox{for $\ell$ even,}
\end{array}\right.\]

\noindent where $d_\ell=F_{\ell-2}$ for even $\ell$ and $d_\ell=F_{\ell-1}+F_{\ell-3}$ for odd $\ell$ (recall that $F_0=1$, $F_1=2$ and $F_k:=F_{k-1}+F_{k-2}$ for every $k\geq 2$).

\end{Th}

(For a proof, as well as for proofs of the other results stated here, see \cite{RV}.)

The identity element of ${\cal G}_\ell^*$ is the element $\widetilde{(01)^\ell}=\widetilde{(10)^\ell}$. We could also identify it with $\widetilde{0^{2\ell}}$, but technical reasons show that it is better to avoid this latter word (hence the star in the notation). The identity element can be seen as the (non-unique) admissible form of $\widetilde{1^{2\ell}}$. This remark can be used to get an algorithm that produces the opposite of a given element $\widetilde{W}$ of ${\cal G}_\ell^*$: in $\widetilde{W}$, replace each $0$ by a $1$ and each $1$ by a $0$, then make admissible the obtained circular word by applying the preceding process to get an admissible circular word: this is the opposite of $\widetilde{W}$.

The sequence of cardinalities of ${\cal G}_\ell^*$ is the integer sequence A004146 of \cite{Sloane}. Its first terms are $1$, $5$, $16$, $45$, $121$, $320$,\ldots We will make use of this observation in section \ref{Graphes}.

As regards admissible circular words of order $q$, we have the following result.

\begin{Th}\label{Ordreq} For $q\geq 1$, let ${\cal P}_q^*$ be the set of circular admissible words $\widetilde{W}$ of even length, containing at least one $1$, and satisfying $q\widetilde{W}=\widetilde{(01)^{|W|/2}}$ (this latter being identified with $\widetilde{(10)^{|W|/2}}$). Assume also the identifications $\widetilde{W}=\widetilde{W^n}$ for any $n$. The set ${\cal P}_q^*$ equipped with the addition 
\[\widetilde{W}\oplus\widetilde{W'}:={\tilde Z}\left(\widetilde{W^{m/|W|}}+\widetilde{W'^{m/|W'|}}\right),\ \mbox{where $m=\lcm(|W|,|W'|)$}\]
\noindent is an abelian group isomorphic to $(\Z/q\Z)\times(\Z/q\Z)$.
\end{Th}

An explicit description of the set ${\cal P}_q^*$ is given by the following theorem. In the sequel, we denote by $N$ the application defined on words as well as circular words by $N(w_0\ldots w_{\ell-1}):=\sum_nw_nF_n$.

\begin{Th}\label{Descr} For any $q\geq 2$, the minimal value $\ell$ for which ${\cal P}_q^*\subseteq{\cal G}_\ell^*$ satisfies the formula
\[2\ell=\min(n\geq 2,\ \mbox{$n$ even\ : } (F_n\bmod q)=(F_{n-1}\bmod q)=1).\]

Moreover, let $\widetilde{\Pi}:=\widetilde{\pi_0\ldots\pi_{2\ell-1}}$ (resp. $\widetilde{\Pi'}:=\widetilde{\pi'_0\ldots\pi'_{2\ell-1}}$) be the circular word of length $2\ell$ such that $N(\Pi)=(F_n-1)/q$ (resp. $N(\Pi')=(F_{n-1}-1)/q$). We have $\widetilde{\sigma(\Pi')}=\widetilde{\Pi}$ and, for any $1\leq i\leq q$:
\[i\cdot\widetilde{\Pi}=\widetilde{i\Pi}\qquad\mbox{and}\qquad i\cdot\widetilde{\Pi'}=\widetilde{i\Pi'}.\]

The circular words $\widetilde{\Pi}$ and $\widetilde{\Pi'}$ are the only non-trivial elements of ${\cal P}_q^*$ satisfying this latter property.
\end{Th}

\section{Periodic expansions}

Circular words are a natural tool for studying numbers with periodic expansion in numeration systems. The first motivation for their study was to give a description of the set of periodic {\em ${\cal F}$-adic numbers}, which are the equivalent of $p$-adic numbers for the Fibonacci sequence $(F_n)_n$. More precisely, an ${\cal F}$-adic number is an infinite admissible word $W=w_0w_1\ldots$ associated with the divergent series $\sum_nw_nF_n$. Assuming the identification $(01)^\infty=(10)^\infty$ and some other identifications derived from this one, the set of ${\cal F}$-adic numbers is an abelian group. Admissible circular words is then the basic tool to get the following characterization of periodic ${\cal F}$-adic numbers.

\begin{Th}\label{F-Rationnels} An ${\cal F}$-adic number $x$ is ultimately periodic iff there exists integers $p$ and $q$ such that $qx=p$. Moreover, the integers $p$ and $q$ being given, the equation $qx=p$ admits exactly $q$ different roots (or $q+1$ if $q$ divides $p$).
\end{Th}

The set of roots of $qx=p$ can be fully described with the help of Theorem \ref{Descr}.

\section{$\gcd$-property of integer sequences}

Let us say that an integer sequence $(u_n)_{n\geq 1}$ has the {\em $\gcd$-property} whenever, for any $m$ and $n$, we have $\gcd(u_m,u_n)=u_{\gcd(m,n)}$. It is well-known that the Fibonacci sequence $(f_n)_n$ defined by $f_1:=f_2:=1$ and $f_n:=f_{n-1}+f_{n-2}$ has the $\gcd$-property (see \cite{HW}).

An immediate consequence of the definition of circular words and of Theorem \ref{Groupe} is that, for any integer $n$, the application $g$: ${\cal G}_\ell^*\longrightarrow{\cal G}_{n\ell}^*$ defined by $g(\widetilde{W}):=\widetilde{W^n}$ is a injective morphism of groups. We can easily deduced from this fact and from Theorem \ref{Groupe} that the sequence $(d_\ell)_\ell$ has the $\gcd$-property. Since $d_{2\ell}=F_{2\ell-2}=f_{2\ell}$, this provides a new partial proof that the Fibonacci sequence $(f_n)_n$ has the $\gcd$-property (limited to even indices).

\section{Balanced partition of the beginning part of the Fibonacci word}

The {\em Fibonacci word} $M=abaababaabaab\ldots$ is the fixed point of the substitution defined by $a\mapsto ab$ and $b\mapsto a$. It is well-known from the theory of sturmian words that $M$ has the {\em balanced property}: denoting by $|W|_a$ the number of $a$s in the word $W$, we have, for any factors $W$ and $W'$ of $M$ such that $|W|=|W'|$, the inequality
\[||W|_a-|W'|_a|\leq 1.\]

One may ask for factors of $M$ of the same length and with exactly the same numbers of $a$s (and, hence, the same numbers of $b$s). The following result gives a answer for factors in the beginning of $M$.

\begin{Th}\label{MotFibo} For any $\ell>2$, let $N_\ell:=bM_{F_{2\ell-2}}$. (Here, we denote by $W_n$ the prefix of $W$ of length $n$.) Define the words $A^{(1)}$,\ldots, $A^{(k)}$ by $N_\ell=A^{(1)}\cdots A^{(k)}a$ and $|A^{(i)}|=d_\ell$, where $k=F_{2\ell-2}/d_\ell$ (yes, $k$ is an integer). The value $|A^{(i)}|_a$ (and, hence, the value $|A^{(i)}|_b$) does not depend on $i\leq k$.
\end{Th}

Theorem \ref{MotFibo} is a quite unexpected application of circular words; its proof involves an analysis of the structure of the set ${\cal G}_\ell^*$. Let $X\in\{(01)^\ell, (10)^\ell, (11)^\ell\}$. For any $\widetilde{W}\in{\cal G}_\ell^*$ different from the identity element, we say that $\tilde{W}$ is {\em of type $X$} whenever we have the equality
\[N(\widetilde{W})+N(-\widetilde{W})=N(X).\]

Such a definition leads to a partition of ${\cal G}_\ell^*$ into three subsets ${\cal T}_X$. (The identity element has to be considered separately: we put $(01)^\ell$ in ${\cal T}_{(01)^\ell}$ and $(10)^\ell$ in ${\cal T}_{(10)^\ell}$; excluding $(01)^\ell$ and $(10)^\ell$ defines ${\cal T}_{(01)^\ell}^*$ and ${\cal T}_{(10)^\ell}^*$.) The partition can be describes in the following way: $\widetilde{W}\in{\cal T}_{(01)^\ell}$ iff $W$ admits $0^{2m}1$ as a prefix and $0$ as a suffix; $\widetilde{W}\in{\cal T}_{(10)^\ell}$ iff $W$ admits $0^{2m+1}1$ as a prefix; $\widetilde{W}\in{\cal T}_{(11)^\ell}$ iff $W$ admits $0^{2m}1$ as a prefix and $1$ as a suffix. This characterization shows that ${\cal T}_{(01)^\ell}=\sigma({\cal T}_{(10)^\ell})$. Morevoer, the sets $N({\cal T}_X)$ have the following relevant properties:

\[N({\cal T}_{(10)^\ell}^*)=\{1+2|M_k|_a+|M_k|_b,\ 0\leq k<F_{2\ell-2}\},\]
\[N({\cal T}_{(01)^\ell}^*)=\{1+3|M_k|_a+2|M_k|_b,\ 0\leq k<F_{2\ell-2}\},\]
\[N({\cal T}_{(11)^\ell}^*)=\{F_{2\ell-1}+3+5|M_k|_a+3|M_k|_b,\ 0\leq k<F_{2\ell-5}-1\}.\]

Now, consider the circular words $\widetilde{\Pi}$ and $\widetilde{\Pi'}$ defined by Theorem \ref{Descr}, for $q:=d_\ell$. It can be proved that, for any $k\leq q$, $\widetilde{k\Pi}$ is of type $(10)^\ell$ and $\widetilde{k\Pi'}$ is of type $(01)^\ell$. We then finally get that, for any $0<i\leq d_\ell$:
\[N(\Pi)=iN(\Pi)-(i-1)N(\Pi)=2|A^{(i)}|_a+|A^{(i)}|_b.\]

The balanced property of $M$, together with the fact that $2|A^{(i)}|_a+|A^{(i)}|_b$ is constant, implies that $|A^{(i)}|_a$ and $|A^{(i)}|_b$ are constant.

\section{Group structure on a set of spanning trees}\label{Graphes}

A graph $G$ being given, recall that a {\em spanning tree} of $G$ is a subgraph of $G$ without cycle and containing all vertices of $G$. For $\ell\geq 1$, the {\em $\ell$-wheel} ${\cal W}_\ell$ is the graph made of $\ell+1$ vertices such that $\ell$ of them are arranged in a cycle and such that the last vertex, the {\em center}, is linked to all the $\ell$ other ones by an edge. The following result can be found in \cite{Truc}:

\begin{Th}\label{Roues} For any $\ell\geq 1$, let $c_\ell$ be the number of spanning trees of the $\ell$-wheel ${\cal W}_\ell$. The sequence $(c_\ell)_\ell$ is the sequence A004146.
\end{Th}

Of course, this strongly suggests a link with the groups ${\cal G}_\ell^*$. A way to make the link more explicit is given by the following

\begin{Prop}\label{Taxonomie} The wheel ${\cal W}_\ell$ being given, let $v_0$ be a vertex of its cycle. A sense of rotation being chosen, write $v_1$, \ldots,  $v_{\ell-1}$ for the successive vertices of the cycle, and let $c$ be the center of the wheel. Denote the edges of ${\cal W}_\ell$ by $r_0:=cv_0$, $s_0:=v_0v_1$, $r_1:=cv_1$, $s_1:=v_1v_2$, and so on until $r_{\ell-1}:=cv_{\ell-1}$ and $s_{\ell-1}:=v_{\ell-1}v_0$. For any spanning tree  ${\cal T}$ of ${\cal W}_\ell$, define the circular word $\widetilde{W}:=\widetilde{w_0\ldots w_{2\ell-1}}$ in the following way:

\[\mbox{$w_{2i}:=$}\left\{\begin{array}{ll}1 & \mbox{if $r_i\in{\cal T}$;}\\
0 & \mbox{otherwise.}\end{array}\right.\qquad \mbox{$w_{2i+1}:=$}\left\{\begin{array}{ll}0 & \mbox{if $s_i\in{\cal T}$;}\\
1 & \mbox{otherwise.}\end{array}\right.\]

The function ${\cal T}\longmapsto {\tilde Z}(\widetilde{W})$ is one-to-one.

\end{Prop}

Hence, such a ``taxonomy function'' equip the set of spanning trees of ${\cal W}_\ell$ with a group structure.

It is interesting to note that the circular words $\widetilde{W}$ defined by the spanning trees of ${\cal W}_\ell$ are not necessarily admissible, but are characterized by the following property: they are exactly the circular words of length $2\ell$ whose blocks of $0$ are of even length (the circular word $\widetilde{0^{2\ell}}$ being excluded). This gives an alternative way of writing the elements of ${\cal G}_\ell^*$, which has the interesting property of unicity (the identity element $\widetilde{(01)^\ell}=\widetilde{(10)^\ell}$ being now written in the unique form $\widetilde{1^{2\ell}}$). In particular, this provides a convenient way to generalize our group structure to circular words of odd lengths.

\section{Other questions}

Here, admissibility was defined by the classical constraint derived from the greedy algorithm applied to the numeration system in base $(1+\sqrt{5})/2$ (or the Zeckendorf numeration system of integers). A natural generalization consists in looking at other constraints of finite type: Tribonacci condition (words without the factor $111$) or, more generally, $k$-bonacci condition (words without the factor $1^k$ for some fixed $k$); words on an alphabet with more than two letters (for example: words on $\{0,1,2\}$ without two successive $2$ - this corresponds to the algebraic equation $2+2X=X^2$), etc. It seems that Theorem \ref{Groupe} can be generalized for many possible definitions of admissibility of finite type (the remark at the end of section \ref{Graphes} also suggests a possible extension to sofic systems). The expected consequences concern the periodic expansion of numbers in other numeration systems, as well as the $\gcd$-property for some classes of linear recurring 
 sequences, and balanced properties of factors of sturmian sequences and generalizations. As regard spanning trees, the point is to find a convenient generalization of wheels (and/or spanning trees).

Another question on circular words is about a multiplicative structure. It is quite concievable that, if a convenient multiplicative structure can be found for admissible words, then things could be said about the expansion of the ${\cal F}$-adic root(s) of $x^2-x-1=0$, since the definition of ${\cal F}$-adic numbers derive from the Fibonacci sequence. One may wonder also whether the sets ${\cal S}_n$ of roots of $F_{n-1}x=F_n$ converge in some sense to what would corresponds to an ${\cal F}$-adic golden ratio.







\bibliographystyle{eptcs}
\bibliography{rittaud}

\end{document}